# Science discussions of retracted articles on Bluesky: public scrutiny or misinformation spreading?

Er-Te Zheng[1], Hui-Zhen Fu[2], Xiaorui Jiang[1], Zhichao Fang[3,4*], Mike Thelwall[1*]

* Corresponding authors

Er-Te Zheng (ORCID: 0000-0001-8759-3643)

[1] School of Information, Journalism and Communication, University of Sheffield, Sheffield, UK.

E-mail: ezheng1@sheffield.ac.uk

Hui-Zhen Fu (ORCID: 0000-0002-1534-9374)

[2] Department of Information Resources Management, Zhejiang University, Hangzhou, China.

E-mail: fuhuizhen@zju.edu.cn

Xiaorui Jiang (ORCID: 0000-0003-4255-5445)

[1] School of Information, Journalism and Communication, University of Sheffield, Sheffield, UK.

E-mail: xiaorui.jiang@sheffield.ac.uk

Zhichao Fang (ORCID: 0000-0002-3802-2227)

[3] School of Information Resource Management, Renmin University of China, Beijing, China.

[4] Centre for Science and Technology Studies (CWTS), Leiden University, Leiden, The Netherlands.

E-mail: z.fang@cwts.leidenuniv.nl

Mike Thelwall (ORCID: 0000-0001-6065-205X)

[1] School of Information, Journalism and Communication, University of Sheffield, Sheffield, UK.

E-mail: m.a.thelwall@sheffield.ac.uk

## Abstract

Post-publication peer review (PPPR) has emerged as an important supplement to traditional peer review, with social media playing a growing role in publicising potential problems in published research. However, it remains unclear whether social media discussions of retracted articles primarily reflect good practices, such as exposing flaws and acknowledging retraction status, or bad practices, such as overlooking retractions and continuing to disseminate scientific misinformation. In this study, we collected Bluesky posts referencing scholarly articles from Altmetric and retrieved metadata for the referenced articles using OpenAlex. The final dataset included 284 retracted articles with 79 pre-retraction posts and 857 post-retraction posts, 59 retraction notices with 186 posts, and 609,461 non-retracted articles with 1,344,756 posts. We manually coded Bluesky posts discussing retracted articles to identify instances of good and bad practice. The results show that posts demonstrating good practice (89.9%) substantially outnumbered those demonstrating bad practice (10.1%). Posts reflecting good practice also had more user engagement. In the pre-retraction phase, good practice posts constituted a slight minority (43.0%), whereas in the post-retraction phase they were dominant (94.2%). Most negative posts in the pre-retraction phase (90.0%) had good practice while only 17.3% positive posts in the post-retraction phase showed bad practice. Thus, sentiment analysis can be helpful to filter posts that could flag potential flaws before retraction, but it may struggle to accurately identify the spread of misinformation after retraction. More broadly, this study highlights the potential of Bluesky to support responsible scientific communication, public scrutiny, and research integrity.

## 1. Introduction

Traditional pre-publication peer review has long served as the primary gatekeeping mechanism for scientific quality, despite its well-recognised limitations (Heesen & Bright, 2021). The growing volume of manuscript submissions (Kusumegi et al., 2025), increasing reviewer fatigue (Breuning et al., 2015; Phuljhele, 2024), and the highly specialised nature of contemporary research have further intensified these limitations. Consequently, studies with methodological flaws, manipulated data, or irreproducible findings may bypass the review process and enter the formal scholarly record (Miske et al., 2026). In response to these challenges, post-publication peer review (PPPR) has gradually emerged as an important supplementary mechanism for scholarly evaluation (Knoepfler, 2015; Peterson, 2018). This development extends the evaluation of research from a closed pre-publication process to an open and continuous form of scrutiny by the broader community after publication. Within this evolving paradigm, social media platforms have become an essential infrastructure for PPPR alongside other sources including the PubPeer commenting site.

Unlike traditional academic channels, which often involve lengthy publication cycles for formal commentaries or letters to the editor, social media enables immediate and decentralised discussion of scientific research (Irawan et al., 2024; O'Sullivan et al., 2021). It provides an open environment where scientists and other readers can share post-publication insights and engage in ongoing dialogue beyond the confines of traditional journals.

### *1.1 Social media discussions of retracted articles*

Retracted articles represent severe instances of scientific failure, and discussions of such articles increasingly take place on social media. Existing studies have shown that retracted articles tend to receive more social media attention than non-retracted articles (Peng et al., 2022; Serghiou et al., 2021). They have also highlighted the potential of these platforms to function as early warning systems for problematic research (Amiri & Sotudeh, 2025; Haunschild & Bornmann, 2021; Zheng, Fu, Thelwall, et al., 2025). Independent scholars and scientific error detectors frequently use social media to publicly flag methodological errors, image duplication, data fabrication, and other concerns, often long before journals or institutions take formal action.

However, the rapid dissemination capacity of social media also introduces substantial risks, revealing the dual role of these platforms in the circulation of flawed science. Many social media users are not researchers and may not follow formal publisher updates, so they may be unaware that a paper has been officially retracted. Additionally, some individuals or groups selectively share retracted research to support pre-existing beliefs, such as anti-vaccine arguments, while disregarding the fact that the findings have been discredited (Durmaz & Hengirmen, 2022). This environment allows flawed research to continue circulating as it were valid science, contributing to the phenomenon of "zombie papers" (Bucci, 2019). Users who are unaware of a paper's retraction status may continue to share, endorse, or cite its invalidated findings. In such cases, social media no longer functions as a mechanism for PPPR, but instead becomes a channel for scientific misinformation.

### *1.2 Scientific discussions on Bluesky*

Historically, X (formerly Twitter) served as the primary platform for studying scholarly communication and PPPR on social media (Haunschild et al., 2021; Haustein, 2019; Holmberg & Thelwall, 2014; Insall, 2023; Mohammadi et al., 2018). However, recent shifts in platform governance, restricted API access, and changes to content moderation algorithms have led to a decline in active academic participation on X (Quelle et al., 2025; Shiffman & Wester, 2025). These developments have encouraged a wider "academic migration" to alternative social media platforms (Vidal Valero, 2023).

In this context, Bluesky has emerged as one of the key alternative platforms for the scientific community (Kupferschmidt, 2024; Mallapaty, 2024). Owing to its accessible interface, high concentration of scholarly information, and decentralised architecture, Bluesky has attracted a growing population of researchers and science communicators (Shiffman & Wester, 2025). For example, discussions of scientific research on Bluesky have been found to show greater originality and sustain higher levels of user engagement than comparable discussions on X (Zheng, Jiang, et al., 2025). These characteristics make Bluesky a particularly suitable environment for contemporary research on scholarly communication through social media.

### *1.3 Sentiment analysis of scientific discussions on social media*

Sentiment analysis is a useful tool for investigating social media posts about research (Hassan et al., 2023; Shahzad & Alhoori, 2022; Thelwall et al., 2011; Walter et al., 2019). In the context of retracted articles, sentiment assumes a distinctive functional role. While positive sentiment in ordinary scientific discourse often indicates approval or endorsement, in the case of flawed research it may reflect the uncritical acceptance and further dissemination of misinformation. Conversely, negative sentiment may represent a corrective response, such as highlighting errors, expressing scepticism, or condemning misconduct.

Earlier studies of social media discussions about retracted articles typically relied on critical keyword dictionaries (Amiri & Sotudeh, 2025; Dambanemuya et al., 2024; Peng et al., 2022) or general-purpose sentiment analysis tools (Amiri et al., 2024; Amiri & Sotudeh, 2025) to identify sentiment. These approaches are operationally straightforward and reasonably transparent, allowing researchers to infer users' attitudes from textual cues. More recently, large language models (LLMs) have been adopted for sentiment analysis and have shown promising performance in textual sentiment classification (Miriyala et al., 2025; Riad et al., 2024; Zhang et al., 2024). In this context, Altmetric has introduced LLM-based sentiment analysis scores. Its system categorises the sentiment of social media posts referencing scholarly articles on a seven-level scale, ranging from *strong negative* to *strong positive*, with the aim of identifying whether a post conveys "a positive, negative, or neutral opinion about the research"[1]. It reportedly achieves 87% accuracy in alignment with human coders, outperforming a previous machine learning approach (Areia et al., 2025). It therefore appears to offer a specialised and suitable tool for analysing sentiment in research-related social media content.

### *1.4 Research objectives and questions*

Although existing studies have examined sentiment in social media discussions of retracted articles (Amiri et al., 2024; Amiri & Sotudeh, 2025), no study has systematically contrasted good and bad practices in such discussions or assessed whether sentiment can help signal these contrasting forms of engagement. To address this gap, this study defines two contrasting roles of social media in relation to retracted articles: good practice and bad practice.

- *Good practice* refers to actions that actively contribute to scientific self-correction. Before a retraction occurs, good practice involves identifying and exposing methodological flaws, fraudulent data, or other problems, thereby acting as an early warning signal for problematic articles (Zheng, Fu, Thelwall, et al., 2025). After a retraction has been issued, good practice involves recognising the retraction and clearly indicating this status when sharing or discussing the research (Serghiou et al., 2021).

---

[1] See more details about the sentiment analysis in Altmetric at: https://help.altmetric.com/en/articles/9807617 (accessed on 11 April, 2026).

- *Bad practice* refers to behaviours that facilitate the spread of scientific misinformation. In the pre-retraction phase, this occurs when users fail to recognise an article's defects and disseminate its flawed findings without cautionary context. In the post-retraction phase, bad practice occurs when users continue to reference or share the article's results without indicating that the article has been retracted (Peng et al., 2022; Serghiou et al., 2021).

By comparing the prevalence of good and bad practice in Bluesky discussions of retracted articles, as well as the engagement they receive, this study assesses whether this emerging platform functions primarily as a mechanism for public scrutiny and PPPR, or whether it risks amplifying erroneous scientific claims. Understanding this balance is essential for evaluating the role of social media in scholarly communication and research integrity.

Specifically, this study addresses the following research questions:

1. How does the distribution of sentiment differ across four types of Bluesky posts: posts referencing retracted articles before formal retraction (pre-retraction posts), posts referencing retracted articles after formal retraction (post-retraction posts), posts referencing retraction notices, and posts referencing non-retracted articles? How do engagement levels vary across posts with different sentiment attributes in these four contexts?
2. What is the proportional distribution of good and bad practices in Bluesky posts referencing retracted articles and retraction notices? How do engagement levels differ between posts reflecting good practice and those reflecting bad practice?

## 2. Data and methods

### *2.1 Data collection of Bluesky posts referencing retracted articles*

The data collection process is summarised in Figure 1. First, we collected all Bluesky posts referencing scholarly articles from the Altmetric database between 1 January 2023 and 30 September 2025. Reposts were excluded, leaving 1,370,952 original Bluesky scholarly posts. We then verified the DOIs of the referenced papers using the OpenAlex database and found that these posts were associated with 611,139 distinct articles.

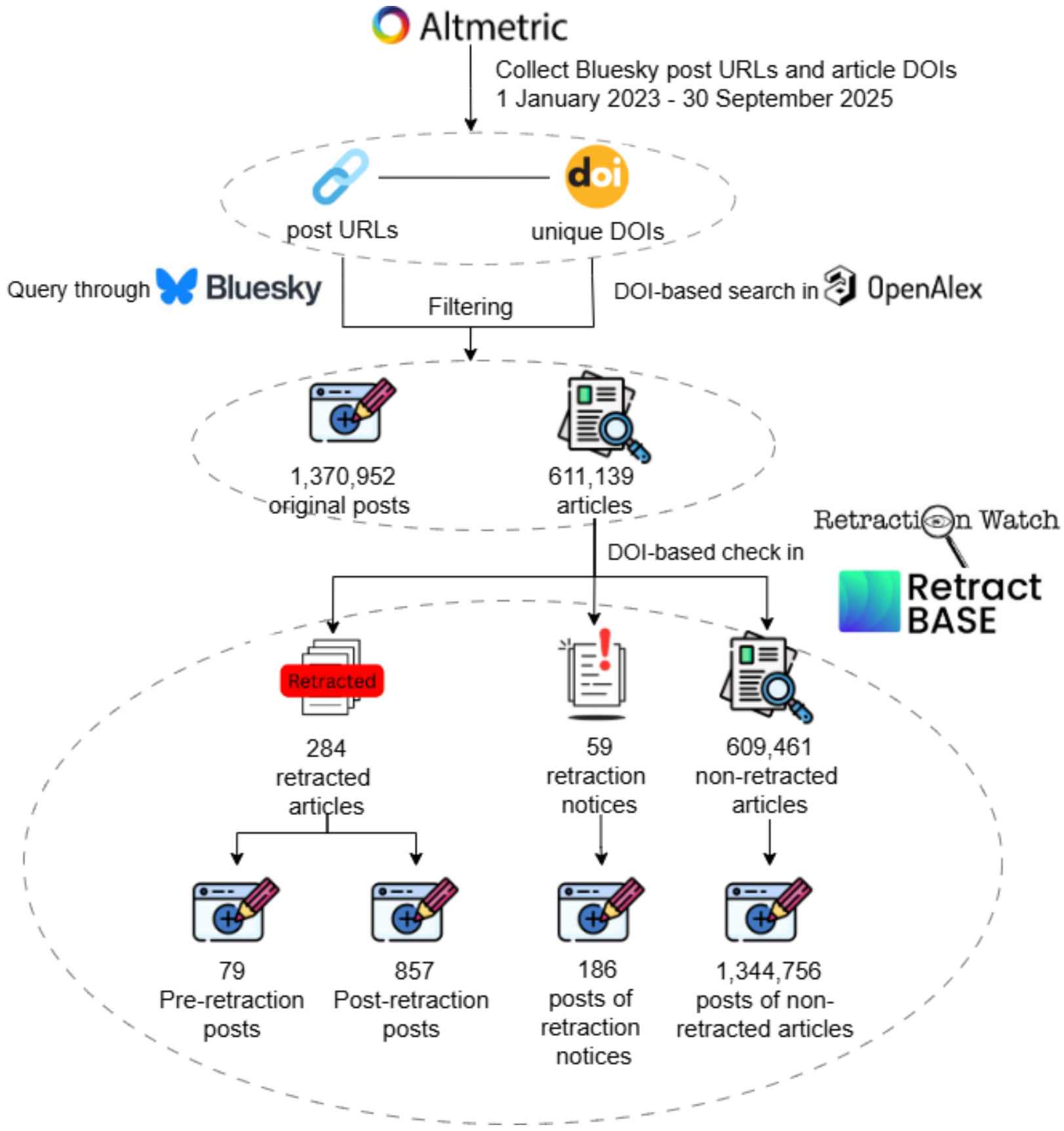

**Figure 1.** Workflow of the post and article data collection process.

We subsequently matched the DOIs of these articles against the Retraction Watch and RetractBASE databases. Retraction Watch appears to be the most comprehensive repository of retracted articles (Brainard, 2018), while RetractBASE fills some gaps by identifying retracted articles in scholarly databases, including PubMed and The Lens (Sánchez et al., 2026). After excluding document types classified as corrections or expressions of concern, the dataset contained 284 retracted articles, 59 retraction notices, and 609,461 non-retracted articles. To establish an accurate timeline, we determined the exact retraction dates by extracting the publication dates of the corresponding retraction notices from OpenAlex. By comparing these retraction dates with the timestamps of the Bluesky posts, we classified each post according to whether it was published before or after the article's retraction.

This procedure allowed us to classify the final dataset into four distinct post types: pre-retraction posts, post-retraction posts, posts referencing retraction notices, and posts referencing non-retracted articles (Table 1). Finally, we collected engagement metrics for each post, including the number of likes, reposts, replies, and quotes, using the Bluesky API.

**Table 1.** Number of posts and referenced articles by post type.

| Post Type | Posts | Articles |
|---|---|---|
| Pre-retraction | 79 | 39 |
| Post-retraction | 857 | 258 |
| Retraction notice | 186 | 59 |
| Non-retracted article | 1,344,756 | 609,461 |

### *2.2 Sentiment analysis of Bluesky posts*

For each Bluesky post in the dataset, we obtained a sentiment score from Altmetric's LLM-based sentiment analysis system (Areia et al., 2025). The seven sentiment categories used by Altmetric are defined in Table 2.

**Table 2.** Sentiment categories of Bluesky posts according to Altmetric.

| Sentiment | Explanation in Altmetric | Example posts |
|---|---|---|
| Strong negative | Expresses strong criticism, warns against, or alerts readers to the mentioned paper. | *Peer-reviewed nonsense. 🙄 Journals are officially self-Sokaling with these GenAI images.* |
| Negative | Casts doubt on, questions, cautions against, or queries the research. | *This is a total fail. The EIC needs to go. Retractions are for fraud and malpractice, not honest mistakes. Science thrives on genuine misfires, they're how we move forward.* |
| Neutral negative | Includes satire, irony, humour, concern, or vague negative cues. | *[Institution] just hit retraction #35. The numbers are getting wild.* |
| Neutral | Contains no sentiment towards the research, such as simply sharing a link or title. | *[Article title] [Link]* |
| Neutral positive | Shares the title and link, possibly with a tag or hashtag, or conveys mild positive framing. | *Fun fact: Blood Type B is basically a 5-star buffet for mosquitoes. Studies show they prefer it over any other type. Pure magnet energy.* |
| Positive | Includes some commentary, suggests reading the research, or uses it to support an argument without explicit praise. | *Game changer: Researchers just proved Einstein's field equation is actually a relativistic quantum mechanical equation. The bridge between gravity and the quantum world is getting real.* |
| Strong positive | Expresses strong recommendation, presents the research as essential or as a solution, or praises it as excellent work or strong evidence. | *Wait, this is actually genius. 🔥 Had no idea this existed until now, but it's exactly what's been missing. Beyond excited for this! 🙌* |

Note: All example posts have been paraphrased to protect user privacy.

### *2.3 Manual coding of good and bad practice*

To classify Bluesky posts as examples of good or bad practice, we employed a mixed-methods content analysis design that integrated both deductive and inductive strategies. All posts referencing retracted articles, comprising 79 pre-retraction posts and 857 post-retraction posts, as well as 186 posts referencing retraction notices, were manually reviewed. In posts referencing retraction notices, users can be assumed to be aware of the article's retracted status, which broadly constitutes good practice. Consequently, this study does not analyse these posts in depth. Instead, they are used as a baseline for comparison when examining user engagement with good and bad practice posts discussing retracted articles.

For posts associated with non-retracted articles, distinguishing between good and bad practices is more difficult. In such cases, it is unclear whether negative critiques reflect mistaken judgement, ideologically motivated attacks on non-existent issues, or reasonable evidence-based questioning. These posts were therefore excluded from the practice classification.

The coding process began with a directed content analysis approach (Hsieh & Shannon, 2005). The first author reviewed the selected posts and classified them as good or bad practice according to the following pre-established operational definitions:

- Good practice refers to identifying and exposing flaws in an article before its retraction, or disseminating the article after its retraction with clear awareness of its retracted status.
- Bad practice refers to failing to recognise an article's flaws before retraction and sharing its findings uncritically as established scientific facts, or circulating the research after retraction without alluding to its retracted status, thus continuing to treat the findings as valid.

Subsequently, an inductive thematic analysis was conducted to identify specific behaviours within these broad categories, following the methodological guidelines of Braun & Clarke (2006). The first author familiarised themselves with the data, generated initial open codes from the text, such as "failing to recognise that the article has been retracted" and "points out problems in the article", and iteratively clustered similar codes into broader thematic categories representing distinct types of comment. The final codebook is presented in Table 3.

To ensure analytical rigour and reduce single-coder bias, we incorporated peer debriefing and collaborative review (Lincoln & Guba, 1985). Specifically, ambiguous posts that did not fit neatly into the predefined categories were documented and discussed within the research team. This iterative and collaborative approach allowed us to resolve discrepancies, reach consensus, and refine the codebook, thereby strengthening the reliability and validity of the final classification matrix.

## 3. Results

### *3.1 Sentiment distribution and user engagement in Bluesky scholarly posts*

The results show that the proportion of negative posts is substantially higher for retracted articles and retraction notices than for non-retracted articles (Figure 2). Posts referencing non-retracted articles provide a baseline sentiment distribution for ordinary science-related discussions on Bluesky, which are predominantly positive (74.2%) and contain relatively little negativity (3.8%). In contrast, sentiment surrounding problematic articles differs markedly from this baseline in the pre-retraction phase. During this period, negative sentiment occurs in 25.3% of posts, while positive sentiment occurs in 41.8% of posts.

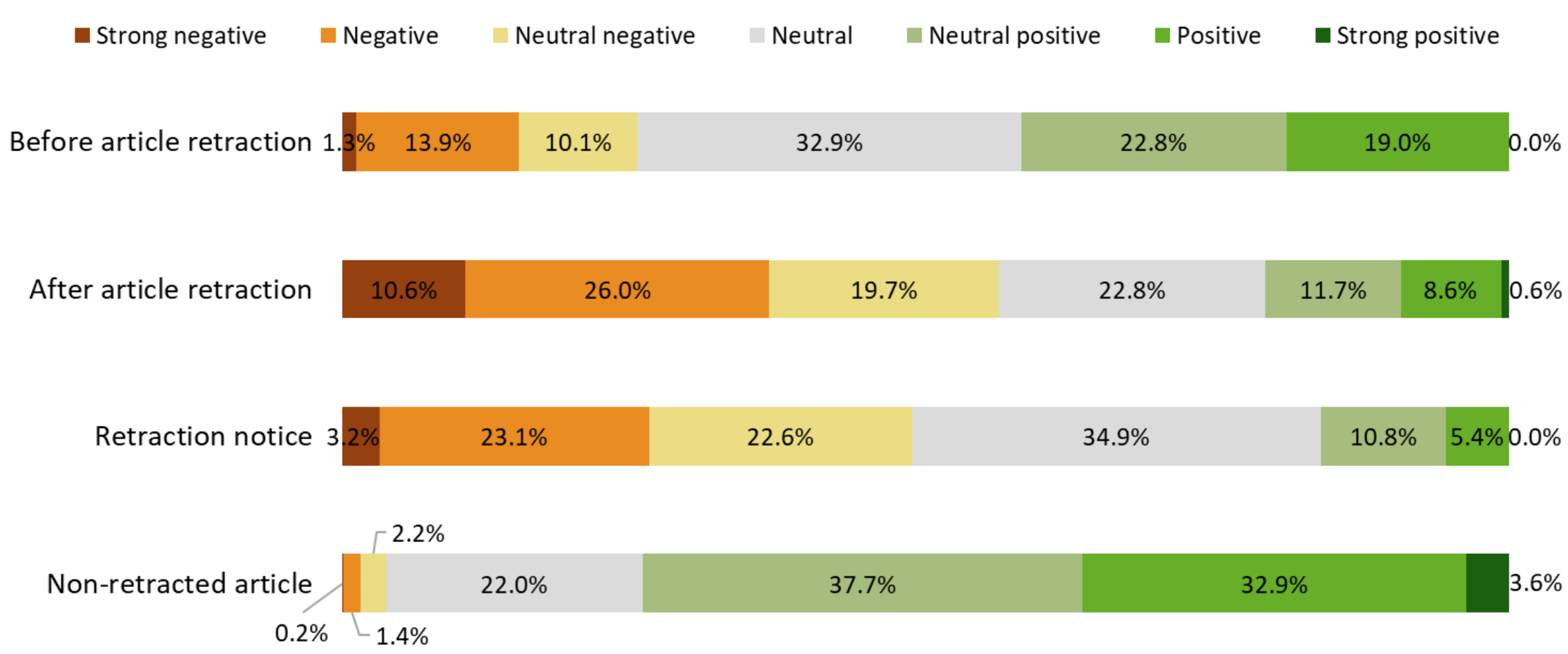

**Figure 2.** Sentiment distribution of Bluesky posts across post types.

Formal retraction amplifies this difference. After article retraction, the proportion of positive posts decreases to about 21%, while the proportion of negative posts more than doubles to 56.3%. This phase also sees an increase in strong negative reactions at 10.6%. However, posts sharing formal retraction notices contain the highest proportion of neutral sentiment and fewer extreme emotional responses. This suggests that users tend to act more as information broadcasters when sharing official announcements, adopting a descriptive tone rather than engaging in direct criticism.

In terms of user engagement, measured by likes, reposts, replies, and quotes, negative posts attract more interactions than almost all positive and neutral posts, even within the control group of non-retracted articles (Figure 3). Posts referencing retraction notices have the highest engagement levels, followed by post-retraction posts. However, pre-retraction posts receive more replies than post-retraction posts, suggesting that problematic articles may provoke debate even before formal retraction. Unlike in other phases, where negative posts consistently drive higher engagement, positive pre-retraction posts have engagement levels comparable to their negative counterparts, particularly in terms of reposts and likes.

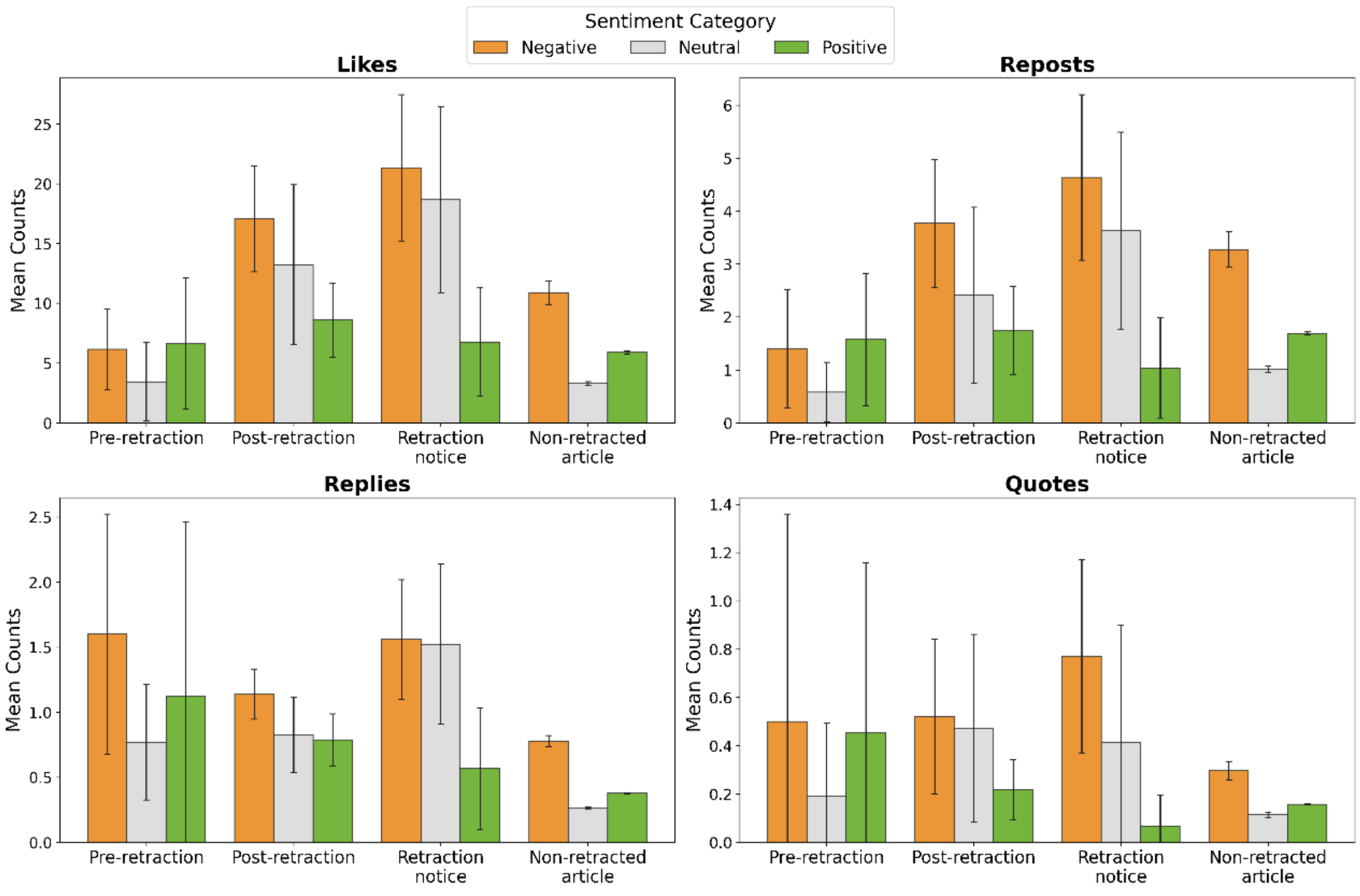

**Figure 3.** Average engagement levels of Bluesky posts across post types. Note: Error bars indicate 95% confidence intervals. Negative sentiment encompasses strong negative, negative, and neutral negative posts. Positive sentiment encompasses strong positive, positive, and neutral positive posts.

### *3.2 Distribution and engagement of good and bad practices in Bluesky scholarly posts*

For a more detailed analysis, we investigated the types of expression contained within each sentiment category for good and bad practices. For this, we manually annotated the content of posts across different sentiment categories and calculated the proportion of each post type (see Figure 4 and Table 3).

In Bluesky discussions of retracted articles, the vast majority of posts demonstrated good practice (89.9%). Before article retraction, just under half of the posts (43.0%) engaged in good practice by identifying problematic research, while the remaining posts contained no warning or criticism. In the post-retraction phase, however, good practice became overwhelmingly dominant (94.2%), as most users explicitly acknowledged the article’s retracted status.

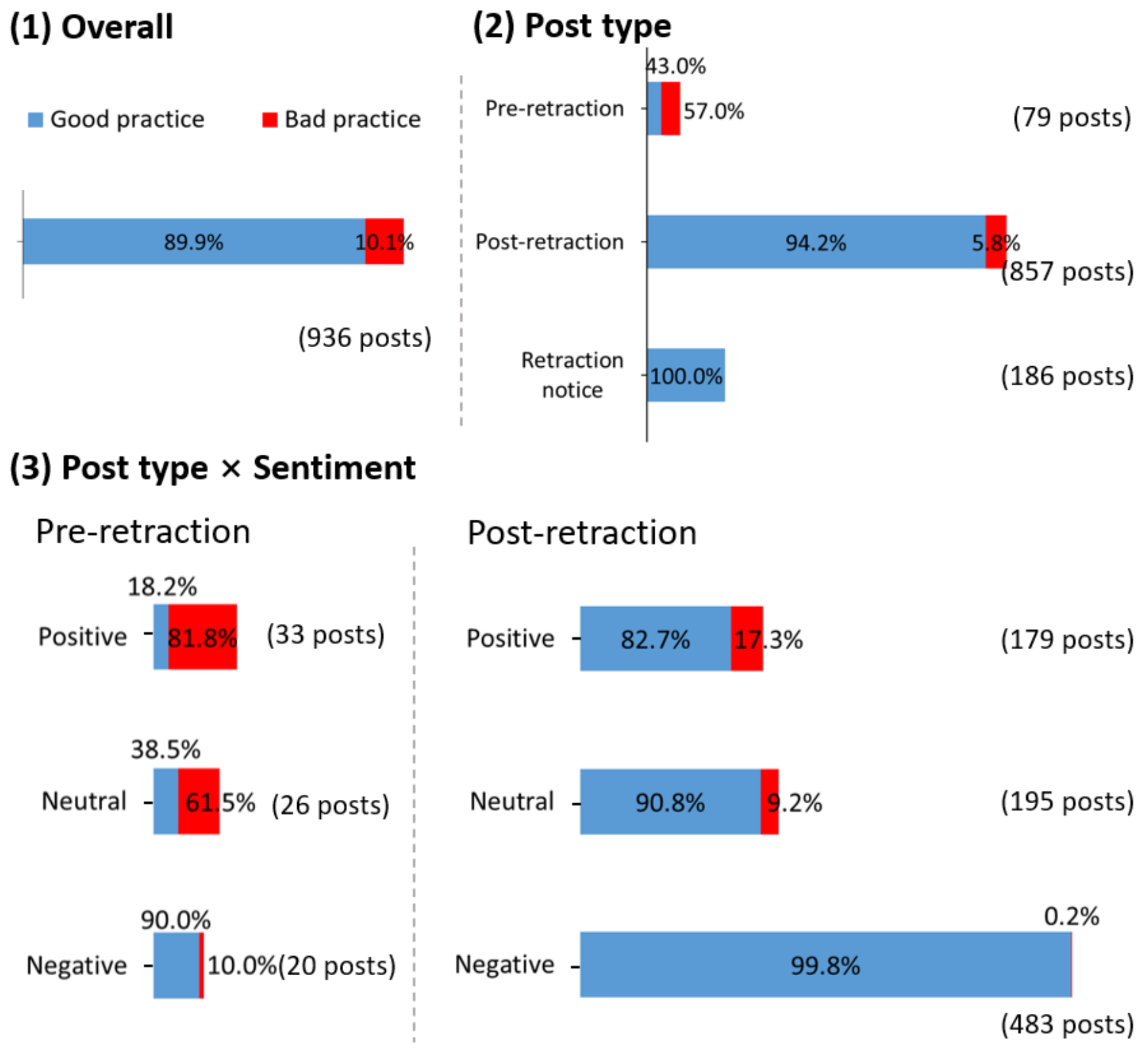

**Figure 4.** Distribution of good and bad practice across different post types.

**Table 3.** Examples and proportions of bad and good practice across post types.

| Post type | Sentiment | Practice type | Comment type | Example post | Proportion within this sentiment (%) |
|---|---|---|---|---|---|
| Pre-retraction | Positive | Bad practice | Praising an article before its retraction | *Get inspired!* | 27/33 (81.8%) |
| | | Good practice | Pointing out problems in the article | *Much of the discourse has centred on the Midjourney-crafted Figure 1, yet Figure 2 remains the standout element of this publication.* | 6/33 (18.2%) |
| | Neutral | Bad practice | Presenting flawed findings as scientific evidence | *You can check out this paper for the facts.* | 16/26 (61.5%) |
| | | Good practice | Pointing out problems in the article | *Elsevier published a paper that accidentally includes an AI's "I am a language model" disclaimer.* | 10/26 (38.5%) |
| | Negative | Bad practice | Presenting flawed findings as scientific evidence | *The healthcare system in this region is on the brink of collapse.* | 2/20 (10.0%) |
| | | Good practice | Pointing out problems in the article | *Peer-reviewed nonsense. 🙄 Journals are officially self-Sokaling with these GenAI images.* | 18/20 (90.0%) |
| Post-retraction | Positive | Bad practice | Failing to recognise that the article has been retracted | *Excited to read this paper!* | 31/179 (17.3%) |
| | | Good practice | Appreciating the retraction itself or thanking those who identified the problem | *A much-needed retraction. Turns out [Institution] spent years conducting unethical and straight-up illegal research on homeless populations. Absolutely wild.* | 148/179 (82.7%) |
| | Neutral | Bad practice | Failing to recognise that the article has been retracted | *Archaeological and genetic data show that the mixing of these human lineages definitely occurred.* | 18/195 (9.2%) |
| | | Good practice | Acknowledging the retraction status or problems in the article | *Retraction: [title].* | 177/195 (90.8%) |
| | Negative | Bad practice | Failing to recognise that the article has been retracted | *Riiiiiiiiiight* | 1/483 (0.2%) |
| | | Good practice | Acknowledging the retraction status or problems in the article | *This paper was packed full of lies.* | 482/483 (99.8%) |

Note: All example posts have been paraphrased to protect user privacy. "Negative" contains "Strong negative/Negative/Neutral negative", "Positive" contains "Strong positive/Positive/Neutral positive".

From the perspective of post sentiment, users engaging in bad practice typically present flawed findings as established facts with positive or neutral overall sentiment. As illustrated in Figure 4, bad practice accounts for a substantial majority of positive posts (81.8%) and more than half of neutral posts (61.5%) before retraction. Users often share these problematic articles with uncritical praise, such as by encouraging others to be inspired by the findings. In contrast, posts reflecting good practice before retraction frequently express negative sentiment. Indeed, most negative posts at this stage represent good practice, using a critical tone to highlight methodological problems before they have been formally recognised through retraction.

After article retraction, negative sentiment is largely used for condemnation, with almost all negative posts representing good practice by acknowledging the retraction status. Most neutral posts also reflect good practice by objectively communicating the article's retracted status. Notably, even positive posts align with good practice in 82.7% of cases. In these instances, users not only recognise the article's retraction status, but also express appreciation for the retraction process and thank those who identified the errors. A small subset of users also highlight the remaining scientific value of the article or commend the authors for exposing unethical reviewers, a phenomenon that also warrants attention.

Despite this broad pattern of community correction, 5.8% of post-retraction posts still show bad practice. These cases mostly appear in positive or neutral posts in which users seem unaware that the article has been retracted (Table 3). This pattern reflects the "zombie paper" phenomenon, in which information lags allow retracted articles to continue circulating as valid science. Nevertheless, such cases are relatively rare in this dataset.

Across all interaction metrics, good practice consistently outperforms bad practice (Figure 5). In the pre-retraction phase, good practice posts attract substantial conversational engagement, particularly in replies and quotes. In the post-retraction phase, the community appears to reward good practice strongly. Posts acknowledging the retraction status of retracted articles or sharing official retraction notices tend to receive the most likes and reposts. Conversely, the small number of bad practice posts in the post-retraction phase receive minimal engagement.

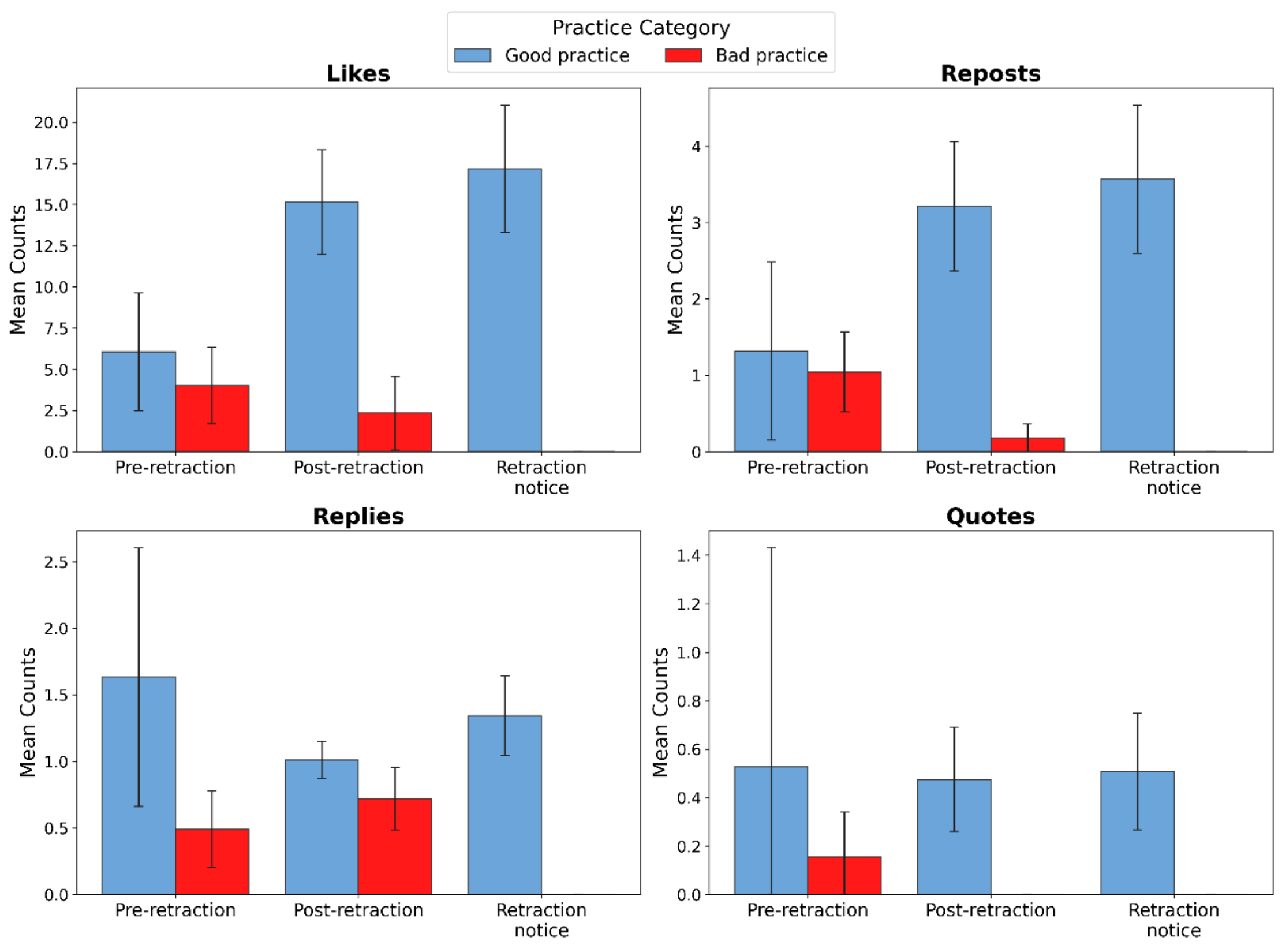

**Figure 5.** Average engagement levels of good and bad practice posts on Bluesky across post types.

## 4. Discussion

### *4.1 Social media discussions of retracted articles on Bluesky*

Overall, Bluesky users tend to discuss retracted articles responsibly. Before retraction, approximately 43% of Bluesky posts were critical of the articles in question. In comparison, on X, only 6% of tweets criticised papers before their retraction (Zheng, Fu, Thelwall, et al., 2025). After retraction, 94.2% of Bluesky scholarly posts pointed out problems with the articles or explicitly acknowledged their retracted status. In contrast, Amiri et al., (2024) reported that neutral sentiment was dominant in discussions of retractions on X. Citation analyses have also shown that only 4% to 6% of post-retraction scholarly citations are made with awareness of the retraction status (Hsiao & Schneider, 2021; Schneider et al., 2020). This may partly reflect publication delays (Hamilton, 2019; Liu et al., 2022), inadequate notification of retraction status on publisher platforms (Bornemann-Cimenti et al., 2016; Elia et al., 2014; Steen, 2011; Suelzer et al., 2021), and the retrieval of static PDFs without retraction watermarks from unofficial repositories (Bar-Ilan & Halevi, 2017; De Cassai et al., 2022). These findings suggest that Bluesky users not only identify potential problems in articles more frequently before retraction than X users, but are also more likely to reference retracted articles with awareness of their retraction status after retraction, particularly when compared with practices observed in traditional scholarly citations.

Regarding engagement metrics, Bluesky posts demonstrating good practice consistently receive higher engagement than those reflecting bad practice across all metrics, including likes, reposts, replies, and quotes. This indicates strong social amplification of corrective information. It also suggests that Bluesky's audience tends to ignore zombie science, thereby reducing its reach. However, before article retraction, the numbers of likes and reposts received by bad practice posts are comparable to those received by good practice posts. This is presumably because users are often unaware of the underlying issues in the articles before formal retraction. In the absence of such awareness, it is understandable that users may uncritically endorse and share the findings. This highlights the continuing risk that Bluesky can still facilitate the spread of scientific misinformation before problems are formally recognised and confirms the importance of formal retractions.

Furthermore, in the post-retraction phase, the number of replies to bad practice posts is nearly as high as that for good practice posts. A review of a sampled subset reveals that these bad practice posts often spark intense debates. For example, other users may mock the original poster for failing to realise the article has been retracted, or engage in arguments over contested topics such as vaccine efficacy. The controversial nature of these bad practice posts therefore appears to drive a higher volume of replies.

The migration of many researchers from X to Bluesky (Arroyo-Machado et al., 2025; Quelle et al., 2025; Shiffman & Wester, 2025), together with evidence that scientific discussions on Bluesky are more original and receive higher levels of engagement (Zheng, Jiang, et al., 2025), may partly explain why discussions on Bluesky appear more responsible than those on X. Bluesky users tend to participate more actively and substantively in discussions of scientific articles, rather than simply reposting article titles, as is common on X (Didegah et al., 2018; Thelwall et al., 2013). As a result, individual posts are more likely to identify problems within articles and provide targeted criticism. On Bluesky, the immediacy of information exchange also enables users to react promptly to retraction events. Such rapid dissemination can amplify awareness of retraction status and reduce the further circulation of misinformation contained in retracted studies.

Overall, discussions of retracted articles on Bluesky suggest a responsive and comparatively responsible environment. Although users sometimes share flawed research before official retractions happen, the platform appears to support subsequent correction. Users often adapt their behaviour after retraction, follow good practices, and amplify corrective information rather than discredited claims. Bluesky therefore appears to provide a useful venue for public scrutiny of scientific research, although it does not eliminate the risk of misinformation before retractions are formally issued.

### *4.2 The utility and limitations of sentiment analysis in social media discussions of retracted articles*

Sentiment analysis provides a useful indicator for social media discussions of retracted articles, as the detected sentiment can suggest whether a post is more likely to reflect good or bad practice. For instance, before retraction, posts with positive or neutral sentiment are more likely to constitute bad practice, whereas posts with negative sentiment tend to represent good practice. Automated sentiment analysis can therefore facilitate the detection of early warning signals for problematic research before retraction (Amiri & Sotudeh, 2025) and help identify the continued dissemination of scientific misinformation after retraction.

In terms of user engagement, negative posts generally receive higher levels of interaction. Posts discussing articles after retraction and posts sharing formal retraction notices also attract more engagement overall. This suggests that user may find negative scientific events and retractions more compelling than ordinary research articles, an observation consistent with Peng et al. (2022). However, before article retraction, negative and positive posts receive almost identical levels of engagement, indicating that many users are likely to be unaware of the article's flaws prior to official action. This highlights the continuing need for caution when interpreting and sharing scientific literature on social media.

Despite its utility and improvements due to large language models, sentiment analysis remains imprecise, and its reliability varies across post types. For example, this study found that many positive posts in the post-retraction phase and posts referencing retraction notices acknowledged the retraction. Rather than positively spreading misinformation, these users expressed appreciation for authors, science sleuths, journals, or other actors involved in correcting scientific errors. Such posts therefore constitute good practice rather than bad practice. Thus, although sentiment analysis offers a rapid and preliminary overview of social media discussions surrounding retracted articles, relying on sentiment alone to classify citation or sharing behaviours may lead to misinterpretation.

For this reason, our study incorporated thematic analysis to achieve a more fine-grained classification of how Bluesky scholarly posts discuss retracted articles. Previous research applying sentiment analysis to social media discussions of retracted articles has similarly employed methods such as thematic coding (Amiri et al., 2024) or thematic clustering (Amiri & Sotudeh, 2025) to refine automated sentiment results. Our results support the effectiveness of this methodological approach.

### *4.3 Implications for the scientific community and the general public*

For the scientific community, the predominance of good practice in discussions of retracted articles suggests that Bluesky may provide a more credible environment than X for scientific communication. It may therefore be a safer platform for typical users, in the sense that they are less likely to encounter misleading scientific information. Furthermore, journals and editors can monitor these decentralised discussions, or use tools such as RetractionRisk Scanner (Zheng, 2025), which aggregates negative social media comments to assess whether published papers may involve reported problems. This proactive approach can support the timely correction or retraction of erroneous and fraudulent literature (Amiri & Sotudeh, 2025; Zheng, Fu, Jiang, et al., 2025). Similarly, researchers can adopt this strategy when assessing papers they read and cite, helping to ensure that the foundational studies on which they rely have not previously been flagged by the community for methodological concerns (Bakker et al., 2024).

For the general public, scientific research has traditionally been perceived as an isolated endeavour conducted within an inaccessible ivory tower. Social media platforms, however, have helped bridge this gap by allowing broader audiences to observe and participate in public discussions of research. By engaging with Bluesky discussions about problematic articles through everyday actions such as liking and reposting, audiences can actively increase the visibility of concerns about flawed research and misinformation (Uppalapati & Nadendla, 2025). This collective amplification may help accelerate the retraction process and support the broader integrity of the scientific enterprise.

### *4.4 Limitations*

This study has several limitations. First, when using Altmetric to collect Bluesky posts referencing scholarly articles, only posts containing explicit article identifiers could be retrieved[2]. Posts that mentioned an article title without including a link, or that displayed the article only within an image, could not be captured during data collection. As a result, some relevant posts were inevitably omitted.

Second, the sentiment of each post was derived directly from Altmetric's sentiment classification. Although this system employs a large-scale sentiment analysis model with a reported accuracy of up to 87% in consistency with human coders (Areia et al., 2025), approximately 13% of posts may still be misclassified, which could influence our results. While we addressed the inherent ambiguity of these broad sentiment categories through manual thematic coding, the baseline automated sentiment data still contains a degree of classification error. Future research would benefit from developing more context-aware and domain-specific sentiment analysis models, such as Aspect-Based Sentiment Analysis (ABSA), which can identify the specific targets of user sentiment (Juroš et al., 2024; Liskowski & Jankowski, 2026; Shukla et al., 2026). This approach would further reduce classification errors in studies of scientific discourse on social media.

[2] https://www.altmetric.com/about-us/our-data/how-does-it-work/.

Third, the overall number of Bluesky posts discussing retracted articles remains relatively small. This is primarily because Bluesky only entered public beta in February 2024, while article retractions typically occur after a considerable time lag. Consequently, the total number of retractions within the study period is limited. This shortage is especially noticeable for pre-retraction posts, as the window for users to identify and discuss flaws before an official retraction was inherently narrow on this newly launched platform. Studies based on such a small sample may yield biased conclusions. Future research using larger retraction datasets is expected to provide more comprehensive and robust findings.

## 5. Conclusions

Bluesky users generally demonstrate responsible practices when discussing retracted articles. Approximately 90% of scholarly posts on the platform reflect good practice in such discussions. Greater attention should therefore be given to the positive role of social media in supporting responsible science communication and maintaining research integrity, particularly by encouraging broader participation in identifying and exposing problematic literature. However, around 10% of posts still reflect bad practice, with users failing to recognise the retraction status of articles and continuing to disseminate erroneous scientific claims from the retracted articles. The potential of social media to function as a vector for misinformation therefore cannot be ignored. Furthermore, posts demonstrating good practice receive substantially higher engagement than those showing bad practice, indicating that Bluesky users generally respond responsibly to scientific content and tend to amplify corrective information. Overall, a balanced perspective on the role of social media is necessary. By leveraging its advantages while mitigating its shortcomings, social media platforms can be more effectively used to facilitate science communication, support post-publication scrutiny, and contribute to research integrity.

## Acknowledgements

Zhichao Fang is financially supported by the National Natural Science Foundation of China (No. 72304274). Er-Te Zheng is financially supported by the GTA scholarship from the School of Information, Journalism and Communication of the University of Sheffield. Mike Thelwall is supported by the Fundação Calouste Gulbenkian European Media and Information Fund (No.

316177). The authors thank Altmetric, Bluesky, OpenAlex, Retraction Watch, and RetractBASE for providing data access for this research.